\documentstyle[aps,psfig]{revtex} 
\begin{document}
\title{High resolution infrared absorption spectra, \\
crystal field, and relaxation processes in CsCdBr$_3$:Pr$^{3+}$.}
\author{M. N. Popova, E. P. Chukalina}
\address{Institute of Spectroscopy, Russian Academy of Sciences,\\
142190 Troitsk, Moscow Region, Russia}
\author{B. Z. Malkin, A. I. Iskhakova}
\address{Kazan State University, 420008 Kazan, Russia}
\author{E. Antic-Fidancev, P. Porcher}
\address{CNRS-UMR 7574 ENSCP, Laboratoire de Chimie Appliquee de
l'Etat Solide, 11,\\
Rue P. et M. Curie, 75231 Paris Cedex 05, France}
\author{J. P. Chaminade}
\address{Institut de Chimie de la Mati$\grave{e}$re
Condens$\acute{e}$e de Bordeaux,\\
ICMCB--CNRS, Ch\^{a}teau Brivazac, Avenue du Docteur Schweitzer,\\
F-33608 Pessac Cedex, France}
\maketitle

\begin{abstract}
High resolution low-temperature absorption spectra of 0.2\,$\%$ Pr$
^{3+}$ doped CsCdBr$_3$ were measured in the spectral region 2000--7000~cm$
^{-1}$. Positions and widths of the crystal field levels within the $^3H_5$,
$^3H_4$, $^3F_2$, and $^3F_3$ multiplets of the Pr$^{3+}$ main center have
been determined. Hyperfine structure of several spectral lines has been
found. Crystal field calculations were carried out in the framework of the
semiphenomenological exchange charge model (ECM). Parameters of the ECM were
determined by fitting to the measured total splittings of the $^3H_4$ and $
^3H_6$ multiplets and to the observed in this work hyperfine splittings of
the crystal field levels. One- and two-phonon relaxation rates were
calculated using the phonon Green's functions of the perfect (CsCdBr$_3$)
and locally perturbed (impurity dimer centers in CsCdBr$_3$:Pr$^{3+}$)
crystal lattice. Comparison with the measured linewidths confirmed an
essential redistribution of the phonon density of states in CsCdBr$_3$
crystals doped with rare-earth ions.
\end{abstract}




\section{Introduction}

Crystals of rare-earth (RE) doped quasi-one-dimensional double bromides
CsCdBr$_3$ are widely studied, mainly, because of their property to
incorporate RE$^{3+}$ ions in pairs, even at low RE concentrations. This
makes them a promising material for up-conversion lasers. The structure of
CsCdBr$_3$ belongs to the $D_{6h}^4$ space group and consists of linear
chains of face-sharing [CdBr$_6$]$^{4-}$ octahedra along the $c$-axis.
Positional symmetry for Cd$^{2+}$ is $D_{3d}$. RE$^{3+}$ ions substitute for
Cd$^{2+}$ forming centers with different mechanisms of charge compensation.
The main center consists of two RE$^{3+}$ ions placed in the chain on each
side of an adjacent cadmium vacancy, [RE$^{3+}$--Cd$^{2+}$ vacancy--RE$
^{3+}$] \cite{R1,R2,R3}. Both RE$^{3+}$ ions in such a center are
equivalent, their positional symmetry lowers from $D_{3d}$ to $C_{3v}$.
Spectra of many other centers with different configurations were reported in
EPR and optical studies of CsCdBr$_3$:RE crystals. In particular, positional
symmetry for RE$^{3+}$ ions in minor axial centers which are supposed to
appear at higher RE concentrations, namely, [RE$^{3+}$--RE$^{3+}$--Cd$
^{2+}$ vacancy] or [RE$^{3+}$-- Cd$^{2+}$ vacancy--Cd$^{2+}$--RE$^{3+}$],
is again $C_{3v}$ but two RE$^{3+}$ positions are not equivalent, while
in [RE$^{3+}$--Cs$^{+}$ vacancy] centers positional symmetry of RE$^{3+}$
ion is $C_s$.

Though the spectra of CsCdBr$_3$:Pr$^{3+}$ were intensively studied before
\cite{R1,R2,R3,R4,R5}, crystal field energies of the lowest excited
multiplets were not investigated in absorption. The positions of crystal
field levels for the main center, as reported by different authors, differ
by 2--14~cm$^{-1}$. Typical spectral resolution used in the mentioned works
was 0.8--2~cm$^{-1}$, while the only high resolution study (it concerned the
$^1D_2$ level) revealed spectral lines as narrow as 0.1~cm$^{-1}$
\cite{R1,R5}. Peculiar doublet line shape was observed for the
$^3H_4 (\Gamma_1) \rightarrow {}^1D_2 (\Gamma_3)$ transition and
explained by the combined effect of unresolved hyperfine structure and
non-axial crystal strains \cite{R5}.

The very first analysis of the crystal field energies of Pr$^{3+}$ in
CsCdBr$_3$ \cite{R1} showed that the crystal field affecting the symmetric
dimer centers consists of a strong cubic and a weak trigonal components and
that the last one is determined mainly by the quadrupolar (B$_2^0$) term.
This conclusion was confirmed in the studies of the crystal field energies
of Tm$ ^{3+}$ and Ho$^{3+}$ symmetric pair centers where data obtained from
both optical and submillimeter EPR investigations were taken into account
\cite {R6,R7}. The sets of crystal field parameters obtained in Refs.
\cite{R2,R8} from the fitting of the simulated energy level schemes to the
measured 40 energy levels of Pr$^{3+}$ are close to one another and provide
small enough root-mean square deviations (22.4--11.1~cm$^{-1}$) between
calculated and observed energy level data. However, some of the crystal
field parameters in CsCdBr$_3$:RE$^{3+}$ presented in the literature
\cite{R6,R7,R8,R9}, in particular B$_6^0$ and B$_6^6$, vary
nonmonotonically with the occupation number of the 4$f$ electron shell.

We have undertaken the low temperature high resolution infrared absorption
study of CsCdBr$_3$:Pr$^{3+}$ with the aims (i) to determine directly
positions of crystal field levels in the region between 2000 and 7000 cm$
^{-1}$ and (ii) to study the line widths and shapes and thus to obtain
information on relaxation processes and hyperfine interactions. We also
performed crystal field calculations for the main center in the framework of
the semiphenomenological exchange charge model (ECM) \cite{R10}, based on
the analysis of the local lattice structure. Parameters of the ECM were
corrected by fitting the calculated hyperfine splittings of crystal field
levels to the experimental data from our measurements. The electron-phonon
interaction effects, in particular, the low temperature relaxation of
excited states of the (Pr$^{3+})_2$-dimer due to one- and
two-phonon transitions between the crystal field sublevels of the $^3H_5,$
$^3H_6$, $^3F_2$ and $^3F_3$ multiplets, are studied in the framework of
the recently derived rigid ion model of lattice dynamics of doped
CsCdBr$_3$:RE crystals \cite{R11}.

\section{Experiment}

CsCdBr$_3$ crystals containing 0.2\,$\%$ of Pr$^{3+}$ were grown by
the method described in Ref.~\cite{R12}. Crystals easily cleave along the
$c$ -axis. We have prepared a 1.3~mm thick sample (sample A) with the
$c$-axis parallel to the cleaved face. The sample was upheld at a
controlled temperature of 5~K in a helium vapor cryostat.
Unpolarized absorption spectra were measured with a BOMEM DA3.002 Fourier
transform spectrometer in the spectral range between 2000 and 7000~cm$^{-1}$
with a resolution from 1.0 to 0.05~cm$^{-1}$.
In Fourier transform spectroscopy, the quantity Res$=1/{\cal L}$,
where ${\cal L}$ is the maximum optical path difference, is indicated
as the resolution. The full width at half hight (FWHH) of the
instrumental function depends on a particular apodization function
used. We used no apodization and, thus, worked with the narrowest
possible instrumental function (FWHH$=0.6$~Res). It is easy to show
that, in this case, the shape of a spectral line is essentially
unchanged provided that FWHH~$\leq$~Res.

Precision of the experimental line positions was 0.05--0.3~cm$^{-1}$,
depending on a particular line. Precise absolute wave number scale is an
intrinsic property of Fourier transform spectroscopy. Line widths were
determined from absorbance spectra calculated with a zero-line taken at a
half of transmitted intensity. In such a way we tried to take into
consideration the fact that the incident light is unpolarized but a majority
of lines is 100\,$\%$ polarized.

The lowest frequency region of the $^3H_4 \rightarrow {}^3F_3$
spectral transition has also been measured with the resolution of
0.005~cm$^{-1}$ using another sample, 3.1~mm thick and oriented
approximately along the $c$-axis (sample B). It allowed almost pure
${\bf k} \| c$ geometry and, consequently, ${\bf E}, {\bf H} \perp c$
polarization of the incident light.

\section{Experimental results}

Fig.~1 shows low temperature transmittance spectra of CsCdBr$_3$:Pr$^{3+}$
(0.2\,$\%$) corresponding to different infrared transitions. The first
excited state of the ground multiplet $^3H_4$ lies at about 170~cm$^{-1}$
\cite{R1,R2,R3,R4} and is not populated at helium temperatures. So, low
temperature absorption spectra display crystal field levels of the excited
multiplets directly. They are listed in Table~1, together with data from
other publications.

As the ground state is the $^3H_4$($^1\Gamma_1$) singlet, excited
$\Gamma_1 $ states manifest themselves in $\pi$-polarization but $\Gamma_3$
states~--- in $\sigma $-polarization, while $\Gamma_2$ states are silent in
electric dipole approximation (see Table~2). Magnetic dipole transitions
are usually very weak, with a possible exception for the $^3H_4
\rightarrow {}^3H_5$ transition that is allowed for the free Pr$^{3+}$
ion.

The symmetries of the $^3F_3$ crystal field levels have been
determined from a comparison between the spectra of the two
differently oriented samples (samples A and B). The assignement of
the 6483.3~cm$^{-1}$ level as a $\Gamma_3$ level is strongly
supported by the well resolved hyperfine structure (hfs) observed in
high resolution spectra (see inset~1 of Fig.~1d and Fig.~2a). The
point is that while $\Gamma_3$ levels exhibit the magnetic hfs,
$\Gamma_1$ and $\Gamma_2$ do not. The
next $\Gamma_3$ level also shows the traces of unresolved hfs,
namely, the 6503.5~cm$^{-1}$ line has a flat bottom (see inset~2 of
Fig.~1d).

Expansion of the 6483.3~cm$^{-1}$ line into six components yields
0.040~cm$^{-1}$ wide Gaussians. The Gaussian shape of hfs components gives
evidence that an inhomogeneous broadening due to random crystal fields
exceeds a homogeneous broadening due to nonradiative transitions to the
nearest $^1\Gamma_2$ level lying at about 10~cm$^{-1}$ lower
\cite{R1,R2} (we did not observe $^3H_4 (^1\Gamma_1) \rightarrow
{}^3F_3 (\Gamma_2$) transitions forbidden as electric dipole ones).
The lineshapes of the 6496.5~cm$^{-1}$ ($\Gamma_1 \rightarrow
\Gamma_1$) and 6503.5~cm$^{-1}$ ($\Gamma_1 \rightarrow \Gamma_3$)
lines reveal homogeneous broadening of 0.17 and 0.31~cm$^{-1}$, respectively.
Such a broadening comes, evidently, from electron-phonon
interaction and implies a noticeable density of phonon states in the region
between 20 and 30~cm$^{-1}$ corresponding to the distances from the
considered levels to lower crystal field levels.

Other examples of level broadening due to electron-phonon interaction can be
found in the $^3H_5$ multiplet which is situated in the region of about
2400~cm$^{-1}$ (see Fig.~1a). The lowest level of this multiplet lying at
about 2234~cm$^{-1}$ was reported to be the $^1\Gamma_2$ level, from
polarized fluorescence measurements \cite{R1}. In that case, a strong very
narrow (with the width of 0.05~cm$^{-1}$) line observed at 2234.8~cm$^{-1}$
corresponds to the magnetic dipole transition $^3H_4(^1\Gamma_1) \rightarrow
{}^3H_5(^1\Gamma_2)$. As we have already mentioned, such a situation is
possible just for the $^3H_4 \rightarrow {}^3H_5$ spectral multiplet.
The next level lies at 26~cm$^{-1}$ from the bottom of the multiplet and,
though suffers from a phonon relaxation, exhibits hfs as it follows
from the shape of 2261.4~cm$^{-1}$ line. This is in accordance with
the previously made
assignments of this level symmetry as $\Gamma_3$. Three closely spaced
levels at 2316.6, 2331.9 and 2347.8~cm$^{-1}$ are broad. Their widths are
3.0, 3.3 and 3.7~cm$^{-1}$ respectively, and it is not possible to determine
their symmetries. Calculations described below show that an overall
hyperfine splittings of $\Gamma_3$ levels do not exceed 0.5~cm$^{-1}$.
Consequently, the level widths of 3--4~cm$^{-1}$ are, mainly, due to rapid
phonon-assisted decay to lower levels. In accordance with the crystal field
calculations (see below), at least one of these three broad absorption lines
is to be assigned to another center. The next level is more narrow again.
According to the shape of the line at 2546.7~cm$^{-1}$ it contains
unresolved hfs and thus corresponds to the $\Gamma_3$ level. The distances
from this level to lower levels exceed the length of the phonon spectrum ($
\sim $ 180~cm$^{-1}$ \cite{R11}). Because of that, one-phonon relaxation is
not possible.

The highest frequency line observed at 2572.1~cm$^{-1}$ could be attributed
to the $^3H_4 (^1\Gamma_1) \rightarrow {}^3H_5(^2\Gamma_2)$ magnetic
dipole transition. But the $^3H_5 (^2\Gamma_2)$ level lying at 26
cm$^{-1}$ above the $^3H_5 (^3\Gamma_3)$ level (2546~cm$^{-1}$) should
be broadened due to one- and two-phonon transitions to the lower crystal
field sublevels, and the width of the respective spectral line should be
about 0.3--0.5~cm$^{-1}$ (see Table~3), not 0.05~cm$^{-1}$ as observed.
Probably, this line belongs to some other Pr$^{3+}$ center. Weak narrow
lines observed at 2153.6 and 2167.3~cm$^{-1}$ (see Fig.~1a) also
have to be assigned to some other centers. If these levels belonged to the
main center, the level at 2234.8~cm$^{-1}$ would be broadened by the
nonradiative transitions to these levels. It is not the case experimentally.

We were able to find only two levels in the $^3F_2$ multiplet and the
lowest sublevel in each of $^3F_4$ and $^3H_6$. The level at 5073~cm$^{-1}$
which lies far (250~cm$^{-1}$) from a lower level is, evidently, broadened
by unresolved hfs and is thus a $\Gamma_3$ level. The measured hyperfine
splittings and widths of hyperfine sublevels found from the experimental
line shapes are presented in Table~3, to compare with the calculated
hyperfine splittings and the estimated one-phonon decay rates (see the next
section).

Many of the observed levels coincide with those of Refs. \cite{R1,R2,R3,R4}
found from selectively excited fluorescence measurements. We did not observe
any additional structure due to nonequivalence of Pr$^{3+}$ positions in
asymmetric pair centers. Thus, our results support the conclusion of Refs.
\cite{R1,R2,R3} that the main center in CsCdBr$_3$:Pr$^{3+}$ is a symmetric
pair [Pr$^{3+}$--Cd$^{2+}$ vacancy--Pr$^{3+}$].

\section{Crystal field calculations}

The energy level pattern of the $^{141}$Pr$^{3+}$ ion (the nuclear spin
$I= 5/2$, 100\,$\%$ abundant, the 4$f^2$ electron shell) in the trigonal
crystal field can be represented by eigenvalues of the effective
parametrized Hamiltonian
\begin{equation}
H=V_{ee}+H_{so}+\Delta E(L,S,J)+H_{cf}+H_{hf},  \label{Eq1}
\end{equation}
where $V_{ee}$ is the two-body Coulomb energy; $H_{so}$ corresponds to the
spin-orbit interaction (we use in the present work $\zeta = 746.2$~cm$^{-1}$
and Slater integrals $F_2 = 304.4$; $F_4 = 45.47$; $F_6 = 4.41$~cm$^{-1}$
\cite{R13}); $H_{cf}$ is the crystal field Hamiltonian
\begin{equation}
H_{cf}=B_2^0O_2^0+B_4^0O_4^0+B_4^3O_4^3+B_6^0O_6^0+B_6^3O_6^3+B_6^6O_6^6
\label{Eq2}
\end{equation}
($O_p^k$ are the Stevens operators \cite{R10}, the quantization axis $z$
coincides with the symmetry axis $C_3$ of the crystal lattice), and $H_{hf}$
is the nuclear energy responsible for the hyperfine structure of the optical
spectra. Because we neglect many minor interactions (three-body terms,
spin-other orbit interaction, two-body correlations in the crystal field
terms and etc.), shifts of the multiplet centers of gravity $\Delta E(L,S,J$
) are introduced to fit the calculated spectrum to the experimental data.
Only projections of the electron-nuclear magnetic and quadrupolar
interactions on multiplet manifolds with fixed orbital ($L$), spin ($S$) and
total ($J$) angular moments were considered:
\begin{eqnarray}
H_{hf}=A(L,S,J){\bf JI}+
\frac{e^2Q(1-\gamma)}{4I(2I-1)}
V_{zz}\left[3I_z^2-I(I+1)\right]-  \nonumber\\
\frac{3e^2Q \langle r^{-3} \rangle \alpha _J}{
4I(2I-1)}\biggl[\frac
13\left(3J_z^2-J(J+1)\right)\left(3I_z^2-I(I+1)\right)+\frac
12\left(J_{+}^2I_{-}^2+J_{-}^2I_{+}^2\right)+  \\ \label{Eq3}
\frac
12\left(J_zJ_{+}+J_{+}J_z\right)\left(I_zI_{-}+I_{-}I_z\right) + \frac
12\left(J_zJ_{-}+J_{-}J_z\right)\left(I_zI_{+}+I_{+}I_z\right)\biggr].
\nonumber
\end{eqnarray}
Here $A(L,S,J)$ is the magnetic dipole hyperfine constant (in the free ion
$A (^3H_4) = 1.093$~GHz \cite{R14}, the corresponding hyperfine constants
for other multiplets are obtained using this value and neglecting the core
electron polarization contributions which do not exceed 1.5\,$\%$
of the 4$f$ electron contributions \cite{R15}), $Q= -5.9\times 10^{-30}$
m$^2$ \cite{R14} is the nuclear quadrupole moment, $\gamma = -70\pm 10$ is
the Sternheimer antishielding factor  \cite{R16}, $\alpha _J$ are reduced
matrix elements of the second rank spherical operators, $ \langle r^p
\rangle $ are moments
of the radial wave function of the 4$f$ electrons calculated
in \cite{R17}, and the electric field gradient at the Pr$^{3+}$ nucleus equals
\begin{equation}
eV_{zz}={\sum\limits_i}eq(i)\frac{3\cos ^2\theta (i)-1}{R(i)^3},  \label{Eq4}
\end{equation}
where $q$($i$) is the charge (in units of the proton charge $e$) of a
lattice ion with spherical coordinates $R$($i$), $\theta $($i$) and $\varphi
$($i$) in the system of coordinates having its origin at the Pr$^{3+}$
nucleus.

As we have already mentioned in the Introduction, the crystal field
parameters obtained earlier from the fitting to the measured energy levels
of Pr$^{3+}$ \cite{R2,R8} and Nd$^{3+}$ \cite{R9} in CsCdBr$_3$, though
providing small enough deviations between the calculated and observed energy
levels, exhibit nonmonotonic variation within the RE$^{3+}$ family (see
Table~4). Moreover, there are large discrepancies between the
measured in this work hfs of the absorption lines and the hfs, calculated
with the crystal field parameters from Ref.~\cite{R8}. In particular, the
6483~cm$^{-1}$ line exhibits well resolved hfs of total width 0.21
cm$^{-1}$, while the calculated width is 0.11~cm$^{-1}$. The formal
fitting procedure employed in  \cite{R2,R8,R9} evidently brings about
overestimated 6-rank parameters of the trigonal crystal field component.

To clear up relations between the crystal field parameters and their origin,
we estimated them in the framework of the exchange charge model \cite{R10}.
The crystal field is represented as a sum ($B_p^k=B_{pq}^k+B_{pS}^k$) of the
electrostatic field of the lattice ions and the exchange charge field
defined by the parameters
\begin{equation}
B_{pq}^k=-e^2K_p^k(1-\sigma _p) \langle r^p \rangle
{\sum\limits_i}q(i)\frac{O_p^k(\theta
(i),\varphi (i))}{R(i)^{p+1}},  \label{Eq5}
\end{equation}
and
\begin{equation}
B_{pS}^k=\frac{2(2p+1)}7e^2K_p^k{\sum\limits_i}(G_sS_s(i)^2+G_\sigma
S_\sigma (i)^2+\gamma _pG_\pi S_\pi (i)^2)\frac{O_p^k(\theta (i),\varphi (i))
}{R(i)},  \label{Eq6}
\end{equation}
respectively. Here $K_p^k$ are numerical factors \cite{R10}, $\sigma _p$ are
the shielding factors, $\gamma _2 = 3/2$, $\gamma _4 =1/3$, $\gamma _6 =
-3/2$, $G_s$ and $G_\sigma = G_\pi = G_p$ are dimensionless parameters
of the model. The expression (6) involves the sum over only the nearest
neighbors (six bromine ions) of the RE$^{3+}$ ion. The overlap integrals
$S_s  =  \langle 4f, m = 0 | 4s \rangle $, $S_\sigma = \langle 4f, m = 0
| 4p, m =0  \rangle$, and $S_\pi = \langle 4f, m = 1 | 4p,
m=1 \rangle$ have been computed using the radial 4$f$~--- wave function of
the Pr$^{3+}$ ion from Ref.~\cite{R17} and the 4$s$, 4$p$ wave functions of
the Br$^{-}$ ion given in \cite{R18}.  Dependencies of the overlap
integrals on the interionic distance $R$ (in Angstroms) are approximated by
the following expressions:  \begin{equation} \begin{array}{l} S_s = 0.09988
\exp(-0.20882 R^{2.17238}); \\ S_\sigma = 0.04610 \exp(-0.03733
R^{3.0662}); \\ S_\pi = 0.39635 \exp(-0.87514 R^{1.30253}).  \end{array}
\label{Eq7}
\end{equation}

Equations (\ref{Eq5}, \ref{Eq6}, \ref{Eq7}) present the crystal field
parameters as explicit functions of relative positions of the impurity
Pr$^{3+}$ ion and the lattice ions. We analyzed distortions in the relaxed
CsCdBr$_3$ lattice containing the substitutional dimer [Pr$^{3+}$--Cd$
^{2+} $ vacancy--Pr$^{3+}$] in the framework of the same quasimolecular
model as the one already used in \cite{R6} to calculate the structure of
symmetric Tm$^{3+}$ dimers. Values of ion displacements from their
equilibrium positions in the perfect lattice were obtained by minimizing the
potential energy of the cluster which involved the nearest 50 ions around
the symmetric dimer center. These ions belong to the first (12 Br$^{-}$),
second (2 Cd$^{2+}$), third (12 Cs$^{+}$), and fourth (24 Br$^{-}$)
coordination shells of the impurity Pr$^{3+}$ ions (see Table~5).

The energy of the interionic interaction is assumed to be a sum of Coulomb
and non-Coulomb terms. The effective ion charges ($q$(Cd$^{2+}$)$= 1.56$,
$q$ (Cs$^{+}$)$= 0.84$, $q$(Br$^{-}$)$= - 0.8$, $q$(Pr$^{3+}$)$= 2.34$) and
the non-Coulomb force constants, corresponding to interactions between the
nearest neighbors, Pr$^{3+}$--Br$^{-}$, Cd$^{2+}$--Br$^{-}$,
Cs$^{+}$--Br$^{-}$, and Br$^{-}$--Br$^{-}$, have been obtained from
the study of the CsCdBr$_3$ lattice dynamics \cite{R11}. According to
results of simulating the relaxation energy minimum with respect to the ion
displacements (see Table~5), the distance between the Pr$^{3+}$ ions
attracted by the Cd$^{2+}$ vacancy diminishes from the value of the lattice
constant $c=0.6722 $~nm \cite{R19} down to 0.5915~nm. This value is very
close to the distance of 0.593~nm between the Gd$^{3+}$ ions \cite{R20} and
0.5943~nm between the Tm$^{3+}$ ions \cite{R6,R21} in the symmetric pair
centers in CsCdBr$_3$, as determined from the EPR spectra. In the perfect
lattice, the first coordination shell of a Cd$^{2+}$ ion has the radius of
0.277~nm; in the symmetric dimer center six Br$^{-}$ ions nearest to the
Cd$^{2+}$ vacancy are forced outwards, but the displacements of the
Pr$^{3+}$ ions along the dimer axis are so large that the corresponding
Pr$^{3+}$--Br$^{-}$ distance decreases down to 0.275~nm. On the contrary,
the terminal triangles of the Br$^{-}$ ions lag behind the impurity ions
and the corresponding interionic distance increases up to 0.290~nm. These
values agree with the distances (0.264 and 0.286~nm) between the Yb$^{3+}$
and the nearest Br$^{-}$ ions in the intrinsic dimer units in
Cs$_3$Yb$_2$Br$_9$ \cite{R22}.

Taking into account the local lattice deformation, we obtained from Eqns.
(\ref{Eq5}, \ref{Eq6}, \ref{Eq7}) the following crystal-field parameters
(values are in~cm$^{-1}$) for Pr $^{3+}$ ion in the symmetric dimer center
(the electrostatic contribution to the quadrupolar crystal field component
$B_{2q}^0=-e^2(1-\sigma _2) \langle r^2 \rangle V_{zz}/4$ was computed
exactly with the Ewald method):
\[
\begin{array}{l} B_2^0=-445.5(1-\sigma
_2)-1.123G_s-2.828G_p; \\ B_4^0=-42.7(1-\sigma _4)-4.52G_s-10.12G_p; \\
B_4^3=1184(1-\sigma _4)+156.84G_s+341.60G_p; \\
B_6^0=1.68(1-\sigma _6)+1.96G_s+0.4395G_p; \\
B_6^3=13.4(1-\sigma _6)+12.56G_s+13.56G_p; \\
B_6^6=24.9(1-\sigma _6)+27.61G_s+9.06G_p.
\end{array}
\]
Values of the model parameters $G_s=5.4$; $G_p=6.1$ have been determined by
comparing the calculated and measured total splittings of the ground state
$^3H_4$ and of the excited multiplet $^3H_6$, the shielding factors have
been fixed at $\sigma _4 = \sigma _6 = 0$, $\sigma _2 =0.89$ in accordance
with the theoretical estimations \cite{R16}. Thus obtained crystal field
parameters (column $a$ of Table~4) were used as starting values
for the fitting procedure constrained by the additional conditions that the
calculated hfs had to agree with the experimental data.

The final values of the crystal field parameters for the Pr$^{3+}$ centers
as compared with the corresponding parameters of Tm$^{3+}$, Ho$^{3+}$ and
Nd$^{3+}$ centers are given in column $b$ of Table~4. Despite the large
trigonal distortion of the nearest bromine octahedron, the main features of
the crystal field splittings are determined by the cubic component of the
crystal field. The $B_6^3$ and $B_6^6$ parameters involve the largest
changes from the initial (calculated) to the final (fitted) values. These
differences are most probably caused by the overestimated displacements of
the six Br$_7$--Br$_{12}$ ions in the basis plane (see Table~5). The
calculated energy levels of the Pr$^{3+}$ ions presented in Table~1 were
obtained by diagonalizing the Hamiltonian (1) in the space of 516
states (terms $^3H$, $^3F$, $^3P$, $^1I$, $^1G$, $^1D$ were taken into
account) with fitted barycenters of the free-ion multiplets. The measured
splittings of the $^3H_4$, $^3H_5,$ $^3H_6$, $^3F_2$, $^3F_3$, $^3F_4$,
$^3P_1$, $^3P_2$, $^1D_2$ multiplets of Pr$^{3+}$ are satisfactorily
described by our final set of crystal-field parameters (column $b$ in Table
4).

The computed hfs of $\Gamma_1$ and $\Gamma_2$ crystal field singlets
consists of three doublets with the total width $\Delta E$ of no more than
0.02~cm$^{-1}$ which can not be resolved in the optical spectra. The
quadrupolar contributions to the hfs are very small ($\sim 10^{-4}$~cm$^{-1}$
), the calculated hfs of $\Gamma_3$ doublets is actually equidistant with
intervals of no more than 0.1 cm$^{-1}.$ An example of the simulated
envelope of the absorption line $^3H_4 (^1\Gamma_1) \rightarrow
{}^3F_3 (^1\Gamma_3),$ where individual transitions between different
electron-nuclear states are presented by Gaussians of 0.04~cm$^{-1}$ width,
is given in Fig.~2b. The observed irregular shape of the hfs of some
absorption lines (see Fig.~2a and Ref.~\cite{R5}) is most probably caused by
crystal fields of low symmetry. We discussed this point in more details in
another publication \cite{R23}.

The most essential difference between results of our calculations and
fitting procedures performed in \cite{R2,R8} is connected with assignments
of symmetry types of some crystal field energy levels. In particular,
symmetries of the third ($^2\Gamma_3$) and fourth ($\Gamma_1$) sublevels
of the $^3H_5$ multiplet and of the second ($^1\Gamma_3$) and third
($\Gamma_2$) sublevels of the ground $^3H_4$ multiplet are inverted in
\cite {R2,R8} (it should be noted that our scheme of crystal field energies
of the ground state agrees with the assignment of Ref.~\cite{R4}).
Additional measurements of polarized spectra are to be carried out to check
our predictions.

\section{Relaxation broadening of the crystal field levels}

Even at the liquid helium temperature the hfs of the most of excited crystal
field sublevels is masked by the spontaneous relaxation broadening. Strong
electron-phonon interaction effects in CsCdBr$_3$:RE$^{3+}$ crystals
originate from the specific density of phonon states which has large maxima
in the low frequency region (20--40~cm$^{-1}$) in the perfect crystal
lattice (see Fig.~3 and Ref.~\cite{R11}). The Hamiltonian of the
electron-phonon interaction, expanded in a power series in ion displacements
from their equilibrium positions, can be written as follows
\begin{eqnarray}
H_{\mbox{\scriptsize el-ph}}={\sum\limits_{s\alpha }}V_\alpha (s)[u_\alpha
(s)-u_\alpha
(\rm RE)]+ \nonumber \\
+ \frac 12{\sum\limits_{s\alpha \beta }}V_{\alpha \beta }(s)[u_\alpha
(s)-u_\alpha (\rm RE)][u_\beta (s)-u_\beta (\rm RE)]...;
\label{Eq8}
\\
V_\alpha (s)={\sum\limits_{pk}}B_{p,\alpha }^k(s)O_p^k, V_{\alpha
\beta }(s)={\sum\limits_{pk}}B_{p,\alpha \beta }^k(s)O_p^k,
\nonumber
\end{eqnarray}
where $[{\bf u}(s)-{\bf u}(\rm RE)]$ is the difference between dynamic
displacements of the ligand ion $s$ and the RE ion, and $B_{p,\alpha }^k(s),$
$B_{p,\alpha \beta }^k(s)$ are the coupling constants. The probability of
the one-phonon transition between the initial $(i)$ and final $(f)$ states
of the RE ion with the energy gap $\hbar \omega _{if} > 0$ can be
represented as
\begin{equation}
W_{if}=\frac 2\hbar {\sum\limits_{s\alpha s^{\prime }\beta }} \langle
f|V_\alpha
(s)|i \rangle \text{Im}g_{\alpha \beta }(ss^{\prime }|\omega _{if})
\langle i|V_\beta
(s^{\prime })|f \rangle (1+n(\omega _{if})),  \label{Eq9}
\end{equation}
where $n(\omega )$ is the phonon occupation number, and $g_{\alpha \beta
}(ss^{\prime }|\omega )$ are spectral representations of the Green's
functions for differences between the ion displacements \cite{R24}. We
performed a calculation of the relaxation rates for all the crystal field
sublevels within the manifolds of $^3H_5$, $^3F_2$ and $^3F_3$
multiplets using the phonon Green's functions of the perfect (CsCdBr$_3$)
and locally perturbed (impurity dimer centers in CsCdBr$_3$:Pr$^{3+}$)
crystal lattices obtained in \cite{R11}. The formation of a dimer leads to a
strong perturbation of the crystal lattice (mass defects in the three
adjacent Cd$^{2+}$ sites and large changes of force constants). As it has
been shown in \cite{R11}, the local spectral density of phonon states
essentially redistributes and several localized modes corresponding to
different representations of the dimer point symmetry group $D_{3d}$ appear
near the boundary of the continuous phonon spectrum of the unperturbed
lattice. As an example of changes in the phonon spectrum, we present in
Fig.~3 spectral densities $D_{\alpha \alpha }(\omega)$ of
displacement-displacement autocorrelation functions
\begin{equation}
\langle u^2_\alpha (R) \rangle = \int D_{\alpha \alpha }(\omega
)d\omega ,  \label{Eq10}
\end{equation}
in the perfect ($R$ = Cd$^{2+}$) and perturbed ($R$ = Pr$^{3+}$) lattices at
zero temperature. The contributions to $D_{xx}(\omega )$ from the localized
modes of $\Gamma_{3g}$ (196.5~cm$^{-1}$) and $\Gamma_{3u}$(190 cm$
^{-1}$) symmetry, and to $D_{zz}(\omega )$ from the localized modes of $
\Gamma_{1g}$ (191.7~cm$^{-1}$) and $\Gamma_{2u}$(191.2~cm$^{-1}$)
symmetry are presented by Lorentzians with the proper weights and the full
width of 2~cm$^{-1}$.

We took into account interactions of Pr$^{3+}$ ion with its nearest
neighbors (six Br$^{-}$ ions) only. Values of coupling constants were
computed in the framework of the ECM with the same parameters as those used
in the crystal field calculations (explicit expressions of the $B_{p,\alpha
}^k(s),$ $B_{p,\alpha \beta }^k(s)$ are given in \cite{R10,R24}). Matrix
elements of electron operators $V_\alpha (s)$ were calculated with the
eigenfunctions of the Hamiltonian (1). Calculated inverse lifetimes $1/{\tau
_i}=W_i={\sum_f}W_{if}$ (the sum is over states $f$ which belong to the same
multiplet as the crystal field level $i$ does, with lower energies) are
given in Table~3. It is seen that despite many simplifying
approximations there is a good correlation between the measured linewidths
and the relaxation rates obtained with the perturbed Green's functions.
Calculations with Green's functions of the perfect lattice gave
overestimated (up to an order of magnitude higher) values of relaxation
rates. Thus, we conclude that the increased Pr$^{3+}$-ligand elastic
interaction, as compared to the Cd$^{2+}$-Br$^{-}$ interaction, and
corresponding enhancement of correlations between displacements of the
impurity RE$^{3+}$ ion and its neighbors strongly suppresses the
electron-phonon coupling.

It should be noted that the measured width of the doublet
$^3\Gamma_3(^3H_5)$ is five times larger than the estimated total hfs
width, though the
one-phonon relaxation broadening of this level is not possible at low
temperatures. We suppose that this level, and two upper levels $^2\Gamma
_2 (^3H_5),$ $^4\Gamma_3 (^3H_5)$ are essentially broadened due to the
two-phonon relaxation. The probability of the spontaneous emission of two
phonons contains three terms which correspond to the first order
contribution from the nonlinear electron-phonon interaction:
\begin{eqnarray}
W_{if}^{(2a)}=\frac 1\pi {\sum\limits_{s\alpha \beta s^{\prime }\gamma
\delta}} \langle f|V_{\alpha \beta }(s)|i \rangle  \langle
i|V_{\gamma \delta }(s^{\prime})|f \rangle \times \nonumber\\
\int \int \text{Im}g_{\alpha \gamma }(ss^{\prime }|\omega _1)\text{Im}
g_{\beta \delta }(ss^{\prime }|\omega _2)\delta (\omega _{if}-\omega
_1-\omega _2)d\omega _1d\omega _2,
\label{Eq11}
\end{eqnarray}
to the second order contribution from the linear electron-phonon interaction:
\begin{eqnarray}
W_{if}^{(2b)}=\frac 2{\pi \hbar ^2}{\sum\limits_{jl}}\int \int \left( \frac{
M_{lf}^{fj}(\omega _1)M_{il}^{ji}(\omega _2)}{(\omega _{ij}-\omega
_1)(\omega _{il}-\omega _1)}+\frac{M_{il}^{fj}(\omega _1)M_{lf}^{ji}(\omega
_2)}{(\omega _{ij}-\omega _1)(\omega _{il}-\omega _2)}\right)\times
\nonumber\\
\delta (\omega_{if}-\omega _1-\omega _2)d\omega _1d\omega _2,
\label{Eq12}
\end{eqnarray}
where
\begin{equation}
M_{lm}^{fj}(\omega _n)=\sum_{ss^{\prime }\alpha \beta } \langle
f|V_\alpha (s)|j \rangle
\text{Im}g_{\alpha \beta }(ss^{\prime }|\omega _n) \langle l|V_\beta (s^{\prime
})|m \rangle ,
\label{Eq13}
\end{equation}
and to the combined action of the first and second order transition
amplitudes:
\begin{eqnarray}
W_{if}^{(2c)}=\frac 2{\pi \hbar }\sum_{s^{\prime }s^{\prime \prime }\gamma
\delta }{\sum\limits_{\alpha \beta s,j}(} \langle f|V_{\alpha \beta
}(s)|i \rangle  \langle i|V_\delta (s^{\prime \prime })|j \rangle
\langle j|V_\gamma (s^{\prime})|f \rangle +c.c.)\times \nonumber\\
\int \int \text{Im}g_{\alpha \gamma }(ss^{\prime }|\omega _1)\text{Im}
g_{\beta \delta }(ss^{\prime \prime }|\omega _2)\frac 1{(\omega _{ij}-\omega
_1)}\delta (\omega _{if}-\omega _1-\omega _2)d\omega _1d\omega _2.
\label{Eq14}
\end{eqnarray}

The calculated sum of two-phonon relaxation rates $\sum_fW_{if}^{(2a)}$
corresponding to spontaneous transitions from the doublet $^3\Gamma_3(^3H_5)
$ to all lower sublevels of the $^3H_5$ multiplet, induced by the nonlinear
terms in Eqn.~(\ref{Eq8}) equals $1.8\times 10^{10}$~s$^{-1}$ and is
almost exactly
cancelled by the contribution $-1.7\times 10^{10}$~s$^{-1}$ from
''crossed'' terms of Eqn.~(\ref{Eq14}). We have obtained the
remarkably larger contribution
to the inverse lifetime of this doublet ($7.1\times 10^{10}$~s$^{-1}$),
that agrees satisfactorily with its measured linewidth, from the second
order terms given in Eqn.~(\ref{Eq13}). It should be noted that
relaxation processes that involve excitations of localized modes of
$\Gamma_{3u}$ and $\Gamma_{3g}$ symmetry dominate yielding more than
a half of the total relaxation rate.

\section{Conclusion}

High resolution (1--0.005~cm$^{-1}$) infrared (2000--7000~cm$^{-1}$)
absorption spectra of CsCdBr$_3$:Pr$^{3+}$ (0.2\,$\%$) were taken
at low temperatures. Crystal field levels of the $^3H_5$, $^3H_6$, $^3F_2$,
and $^3F_3$ multiplets for the symmetric pair center [Pr$^{3+}$--Cd$
^{2+}$ vacancy--Pr$^{3+}$] were found directly in absorption. Hyperfine
structure, inhomogeneous and relaxation-induced widths of several crystal
field levels were measured.

We performed crystal field calculations in the framework of the
semiphenomenological exchange charge model and took into account local
lattice deformation around an impurity center. Only two parameters of the
model had to be determined by a comparison of calculated and experimental
energy levels. Thus obtained crystal field parameters were corrected by
fitting the calculated hyperfine structure to the measured one. It follows
from our results that despite a large trigonal distortion of the bromine
octahedra in the nearest surroundings of impurity Pr$^{3+}$ ions, crystal
field splittings are determined, mainly, by the cubic component of the
crystal field.

Using the results of the crystal field calculations and of the lattice
dynamics analysis we were able, for the first time, to calculate exactly
(without any additional fitting parameters) one- and two-phonon spontaneous
relaxation rates. The calculated total relaxation rates reveal the
linewidths which are in reasonable agreement with the measured ones. The
obtained information on relative efficiency of linear and nonlinear terms in
the Hamiltonian of the electron-phonon interaction in stimulation of the
phonon emission is important for the thorough derivation of the theory of
multiphonon relaxation in rare-earth compounds.

The peculiarities of the electron-phonon interaction effects in
CsCdBr$_3$:Pr$^{3+}$ crystals originate from the specific density of phonon
states that extends by only $\sim 180$~cm$^{-1}$ in the perfect CsCdBr$_3$
lattice, and from the localized modes induced by the impurity dimer
centers. Another important feature and, probably, the most interesting
result of this study, is a strong suppression of the effective
electron-phonon coupling due to a local increase of elastic forces in the
activated crystal and the corresponding enhancement of correlations between
displacements of the impurity RE ion and its neighbors.

\section*{Acknowledgements}
An encouraging interest of G. N. Zhizhin is acknowledged.  This work was
supported in part by the RFBR (Grant N 99--02 16881) and the Russian
Ministry of Science (Grant N 08.02.28 of the Program "Fundamental
Spectroscopy").


\section*{Figures}

\begin{figure}[h]
\centerline{\psfig{figure=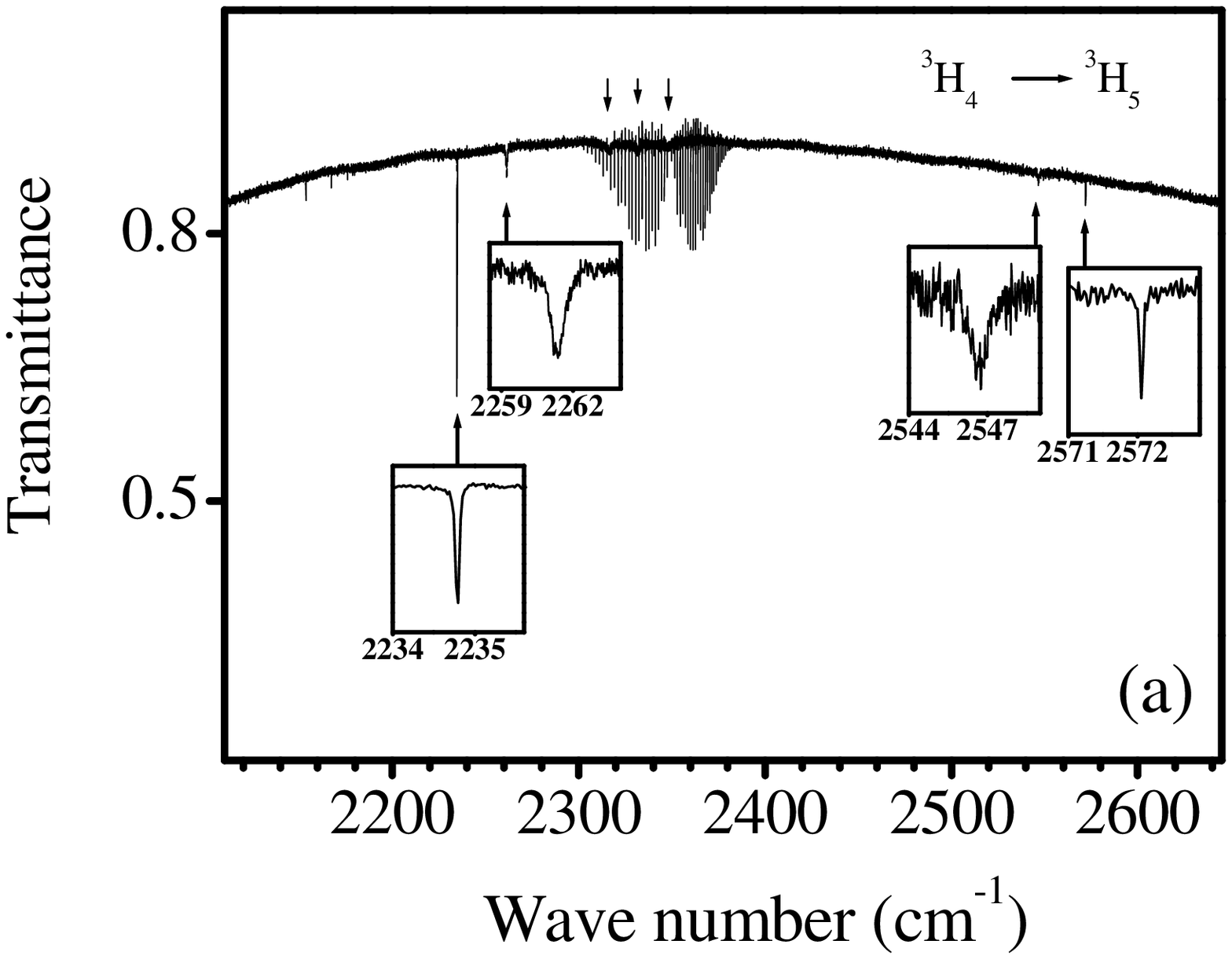,height=200mm}}
\centerline{\psfig{figure=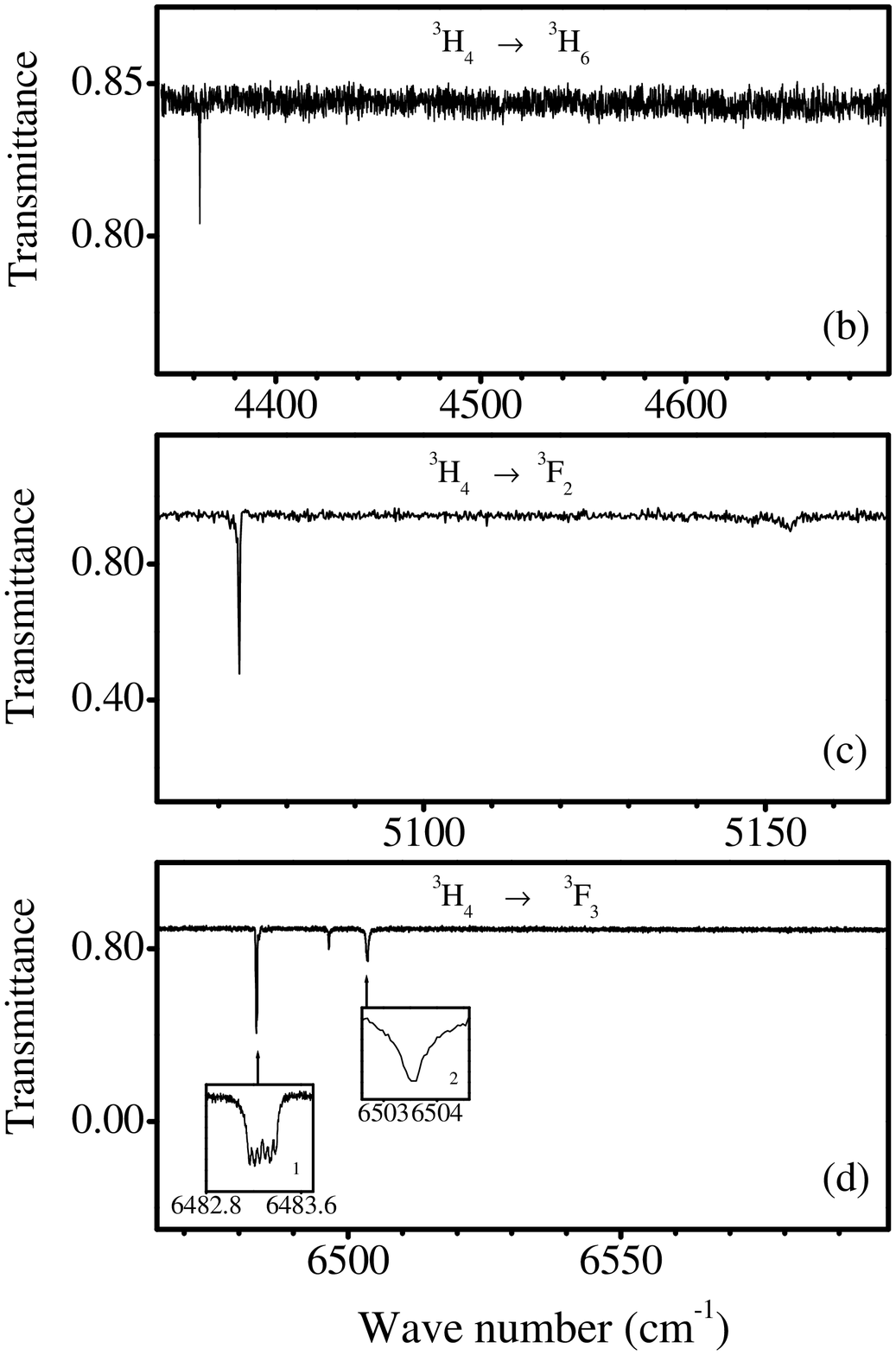,height=200mm}}
\caption{Transmittance spectra of CsCdBr$_3$:Pr$^{3+}$ at 5\,K
corresponding to the optical transitions from the ground state
$^3$H$_4$($^1\Gamma_1$) to the excited (a) $^3$H$_5$, (b) $^3$H$_6$, (c)
$^3$F$_2$, and (d) $^3$F$_3$ crystal field multiplets. Spectral resolution
is 0.055, 0.3, 0.3, and 0.05~cm$^{-1}$ for (a), (b), (c), and (d),
respectively, and 0.005~cm$^{-1}$ for insert~1 of Fig.~1(d).
Very sharp lines in the spectral region
2300--2400~cm$^{-1}$ are due to residual CO$_2$ in the spectrometer.}
\end{figure}

\begin{figure}[h]
\centerline{\psfig{figure=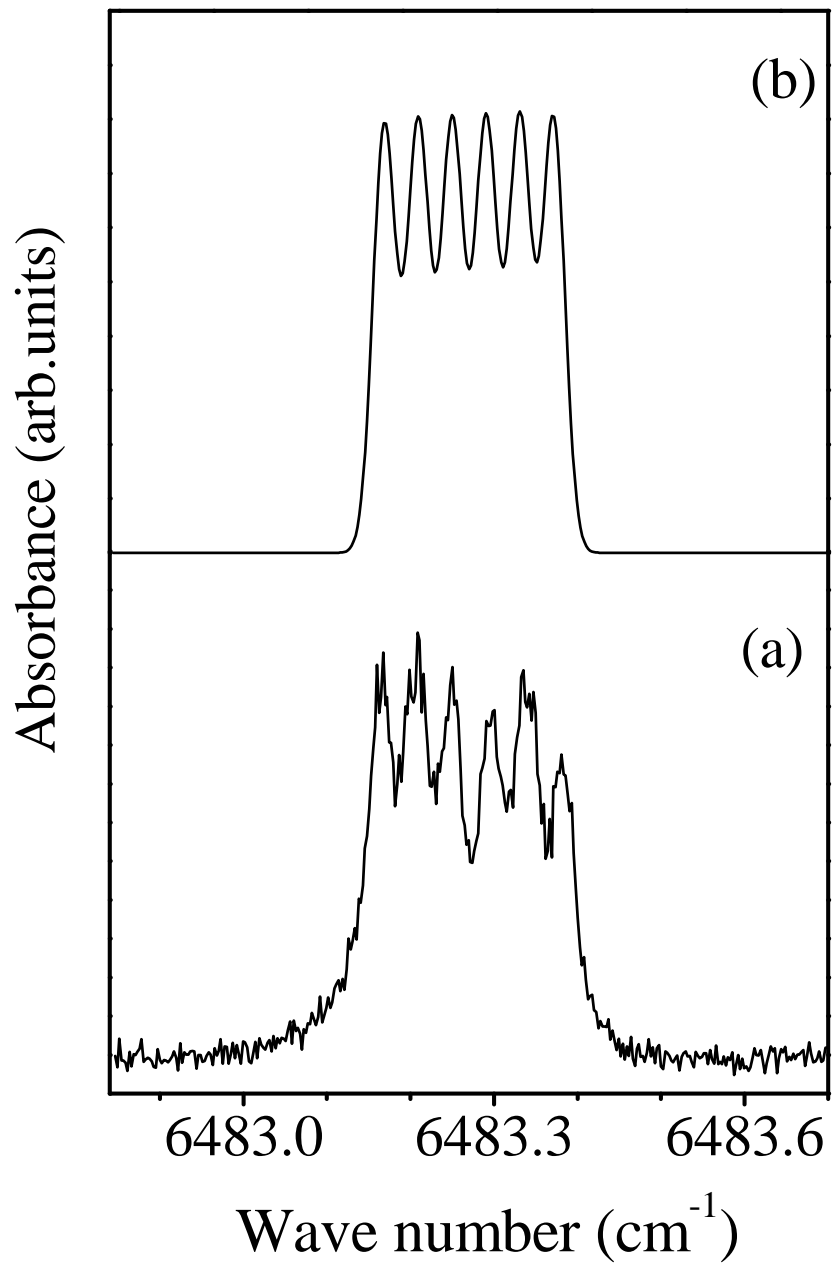,height=200mm}}
\caption{(a) Measured with resolution 0.005~cm$^{-1}$ and (b)
calculated hyperfine structure of the $^3H_4 (^1\Gamma_1) \rightarrow
^3F_3 (^1\Gamma_3)$ transition of CsCdBr$_3$:Pr$^{3+}$.}
\end{figure}

\begin{figure}[h]
\centerline{\psfig{figure=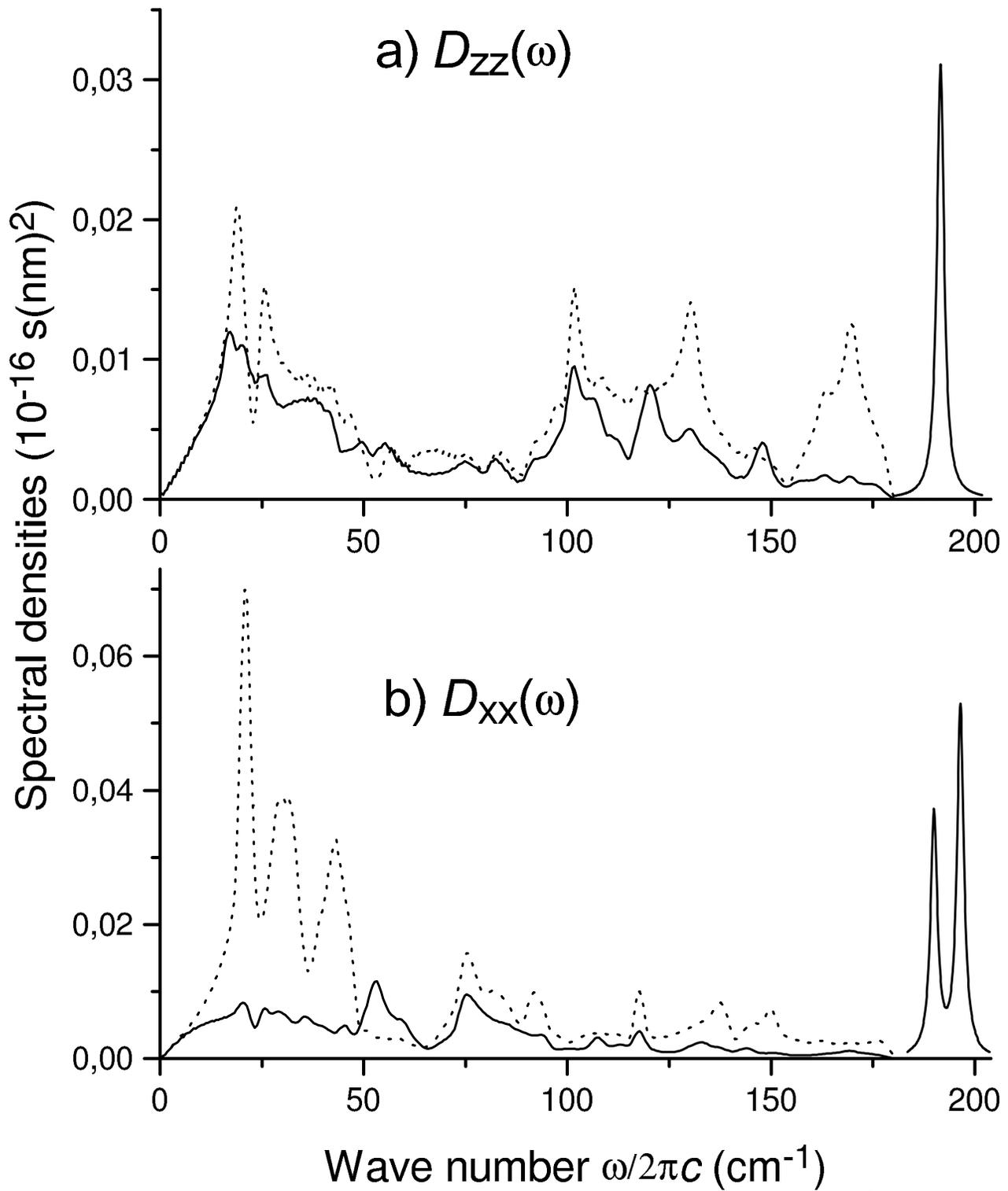,height=200mm}}
\caption{Simulated spectral densities of displacement-displacement
autocorrelation functions for Cd$^{2+}$ in CsCdBr$_3$ (dotted curves) and
for Pr$^{3+}$ in the impurity dimer centers (solid curves).}
\end{figure}


\begin{table}[h]
\caption{Positions (cm$^{-1}$) and symmetries of energy levels for the
symmetric dimer centers, and widths (in brackets,~cm$^{-1}$) of absorption
lines from the ground state in CsCdBr$_3$:Pr$^{3+}$.}
\begin{tabular}{cllllll}
  & \multicolumn{4}{c}{Measured} &  &  \\
Multiplet & \multicolumn{2}{c}{This work} & \multicolumn{2}{c}{Literature
data [1-4]} & \multicolumn{2}{c}{Calculated} \\ \hline
$^3H_4$ & 0 &  & 0 & $\Gamma_1$ & 0 & $^1\Gamma_1$ \\
&  &  & 166--170 & $\Gamma_2$, $\Gamma_3$ ? & 168 & $^1\Gamma_3$ \\
&  &  & 194--196 & $\Gamma_3$, $\Gamma_2$ ? & 188 & $\Gamma_2$ \\
&  &  & 323--328 & $\Gamma_3$ & 293 & $^2\Gamma_3$ \\
&  &  & 551--555 & $\Gamma_1$ & 545 & $^2\Gamma_1$ \\
&  &  & 574--580 & $\Gamma_3$ & 564 & $^3\Gamma_3$ \\
$^3H_5$ & 2234.8 (0.05) & $\Gamma_1$, $\Gamma_2$? & 2230--2236 & $\Gamma_2$
& 2235 & $^1\Gamma_2$ \\
& 2261.4 (0.64) & $\Gamma_3$ & 2256--2264 & $\Gamma_3$ & 2251 & $^1\Gamma_3$
\\
& 2316.5 (3.0) & $\Gamma_1$, $\Gamma_3$? & 2309--2318 & $\Gamma_1$,
$\Gamma_3$? & 2330 & $^2\Gamma_3 $ \\
& 2331.9 (3.3) & $\Gamma_3$, $\Gamma_1$? & 2327--2334 & $\Gamma_3$,
$\Gamma_1$? & 2334 & $\Gamma_1$ \\
& 2348.4 (3.7) &  & 2347 &  &  &  \\
& 2546.7 (0.7) & $\Gamma_3$ & 2533--2547 & $\Gamma_3$ & 2540 & $^3\Gamma_3$
\\
&  &  & 2544--2595 & $\Gamma_2$ & 2593 & $^2\Gamma_2$ \\
&  &  &  &  & 2620 & $^4\Gamma_3$ \\
$^3H_6$ & 4362.8 (0.19) & $\Gamma_3$  & 4362--4364 & $\Gamma_3$ &
4363 & $^1\Gamma_3 $ \\
&  &  & 4374--4378 & $\Gamma_1$ & 4390 & $^1\Gamma_1$ \\
&  &  & 4413--4415 & $\Gamma_3$ & 4414 & $^2\Gamma_3$ \\
&  &  & 4494 & $\Gamma_2$ & 4539 & $^1\Gamma_2$ \\
&  &  & 4700--4704 & $\Gamma_3$ & 4710 & $^3\Gamma_3$ \\
&  &  & 4725--4730 & $\Gamma_1$ & 4733 & $^2\Gamma_1$ \\
&  &  & 4761 & $\Gamma_3$ & 4765 & $^4\Gamma_3$ \\
&  &  & 4781--4808 & $\Gamma_2$ & 4804 & $^2\Gamma_2$ \\
&  &  & 4808--4843 & $\Gamma_1$ & 4815 & $^3\Gamma_1$ \\
$^3F_2$ & 5073.0 (0.11) & $\Gamma_3$ & 5070--5076 & $\Gamma_3$ & 5073 & $
^1\Gamma_3$ \\
&  &  & 5145--5153 & $\Gamma_1$, $\Gamma_3$ ? & 5128 & $^2\Gamma_3$ \\
& 5153.5 (2.8) &  & 5152--5158 & $\Gamma_1$ & 5136 & $\Gamma_1$ \\
$^3F_3$ &  &  & 6473--6479 & $\Gamma_2$ & 6473 & $^1\Gamma_2$ \\
& 6483.3 (0.26) & $\Gamma_3$ & 6480--6488 & $\Gamma_3$ & 6483 & $^1\Gamma_3$
\\
& 6496.5 (0.17) & $\Gamma_1$ & 6499--6500 & $\Gamma_1$ & 6485
& $\Gamma_1$ \\
& 6503.5 (0.4) & $\Gamma_3$  & 6500--6510 & $\Gamma_3$ & 6495 &
$^2\Gamma_3$ \\
&  &  & 6526--6535 & $\Gamma_2$ & 6533 & $^2\Gamma_2$ \\
$^3F_4$ & 6858.0 ($<$1) &  & 6858--6860 & $\Gamma_1$ & 6858 & $^1\Gamma_1 $
\\
&  &  & 6894 & $\Gamma_2$ & 6907 & $\Gamma_2$ \\
&  &  & 6903--6908 & $\Gamma_3$ & 6902 & $^1\Gamma_3$ \\
&  &  & 6914 & $\Gamma_3$ & 6930 & $^2\Gamma_3$ \\
&  &  & 7103--7104 & $\Gamma_1$ & 7115 & $^2\Gamma_1$ \\
&  &  & 7111--7117 & $\Gamma_3$ & 7108 & $^3\Gamma_3$ \\
$^1G_4$ &  &  & 9777 &  & 9777 & $^1\Gamma_1$ \\
&  &  &  &  & 9869 & $^1\Gamma_3$ \\
&  &  &  &  & 9899 & $\Gamma_2$ \\
&  &  &  &  & 9955 & $^2\Gamma_3$ \\
&  &  &  &  & 10427 & $^2\Gamma_1$ \\
&  &  &  &  & 10453 & $^3\Gamma_3$ \\
$^1D_2$ &  &  & 16536--16540 & $\Gamma_3$ & 16536 & $^1\Gamma_3$ \\
&  &  & 16567--16570 & $\Gamma_1$ & 16588 & $\Gamma_1$ \\
&  &  & 17004--17011 & $\Gamma_3$ & 17001 & $^2\Gamma_3$ \\
$^3P_0$ &  &  & 20386--20393 & $\Gamma_1$ &  &  \\
$^3P_1$ &  &  & 20956--20964 & $\Gamma_2$ & 20964 & $\Gamma_2$ \\
&  &  & 21011--21047 & $\Gamma_3$ & 21010 & $\Gamma_3$ \\
$^3P_2$ &  &  & 22106--22118 & $\Gamma_3$ & 22117 & $^1\Gamma_3$ \\
&  &  & 22163--22177 & $\Gamma_1$ & 22165 & $\Gamma_1$ \\
&  &  & 22216--22229 & $\Gamma_3$ & 22223 & $^2\Gamma_3$ \\
&  &  &  &  &  &
\end{tabular}
\end{table}


\begin{table}[h]
\caption{Selection rules for electric dipole (d) and magnetic dipole ($\mu$)
transitions in C$_{3v}$ point group.}
\begin{tabular}{l|lll}
C$_{3v}$ & $\Gamma_1$ & $\Gamma_2$ & $\Gamma_3$ \\ \hline
$\Gamma_1$ & d$_z$ & $\mu_z$ & d$_x$, d$_y$; $\mu_x$, $\mu_y$ \\
$\Gamma_2$ & $\mu_z$ & d$_z$ & d$_x$, d$_y$; $\mu_x$, $\mu_y$ \\
$\Gamma_3$ & d$_x$, d$_y$; $\mu_x$, $\mu_y$ & d$_x$, d$_y$; $\mu_x$, $\mu_y$
& d$_z$; d$_x$, d$_y$; $\mu_x$, $\mu_y$ \\
&  &  &
\end{tabular}
\end{table}

\begin{table}[h]
\caption{The calculated and measured total hyperfine structure widths $
\Delta $E/$\hbar$, the one-phonon relaxation rates W and the widths $
\Delta\omega$ (10$^{10}$~s$^{-1}$) of the crystal field sublevels at the
temperature of 5~K.}
\begin{tabular}{clllll}
$^{2S+1}$L$_J$ & Crystal field & \multicolumn{2}{c}{$\Delta$E/$\hbar$} & W$
^{(a)}$ & $\Delta\omega$ \\
& energy (cm$^{-1}$) & Calculated & Measured & Calculated & Measured \\
\hline
$^3H_5$ & $^1\Gamma_2$ 0 (2235) & 0.05 & $<$0.94 & 0 & 0.94$^{(b)}$ \\
& $^1\Gamma_3$ 26 & 6.68 & 6.41 & 15.5 [113] & 5.65 \\
& $^2\Gamma_3$ 82 & 5.43 &  & 255 [573] & 56.6 \\
& $^1\Gamma_1$ 97 & 0.16 &  & 157 [745] & 62.2 \\
& $^3\Gamma_3$ 312 & 2.65 &  & 7.2$^{(c)}$ & 13.2 \\
& $^2\Gamma_2$ 358 & 0 &  & 52 [59] &  \\
& $^4\Gamma_3$ 385 & 9.03 &  & 59 [311] &  \\
$^3H_6$ & $^1\Gamma_3$ 0 (4363) & 2.00 & 2.26 & 0 & 1.30$^{(b)}$ \\
$^3F_2$ & $^1\Gamma_3$ 0 (5073) & 1.15 & 1.32 & 0 & 0.94$^{(b)}$ \\
& $^2\Gamma_3$ 75 & 2.09 &  & 14 [31] &  \\
& $^1\Gamma_1$ 80 & 0.03 &  & 32 [57] & 52.8 \\
$^3F_3$ & $^1\Gamma_2$ 0 (6473) & 0.03 &  & 0 &  \\
& $^1\Gamma_3$ 10 & 3.71 & 3.96 & 0.52 [9.0] & 0.75$^{(b)}$ \\
& $^1\Gamma_1$ 23 & 0 &  & 4.5 [26] & 3.20 \\
& $^2\Gamma_3$ 30 & 1.53 & 1.69 & 15.8 [95] & 5.85 \\
& $^2\Gamma_2$ 53 & 0 &  & 63 [98] &  \\
&  &  &  &  &
\end{tabular}
$^{(a)}$ - numbers in square brackets were obtained with the phonon spectrum
of the unperturbed crystal lattice, \\$^{(b)}$ - the inhomogeneous width, \\$
^{(c)}$ - two-phonon relaxation rate.
\end{table}


\begin{table}[h]
\caption{Crystal field parameters B$^k_p$ (cm$^{-1}$) of symmetric dimer
centers in CsCdBr$_3$:RE$^{3+}$ crystals.}
\begin{center}
\begin{tabular}{ccllllll}
p & k & Tm (4$f^{12}$) & Ho (4$f^{10}$) & Nd (4$f^3$) & \multicolumn{2}{c}{
Pr (4$f^2$)} &  \\
&  & [6] & [7] & [9] & [8] & \multicolumn{2}{c}{this work} \\
&  &  &  &  &  & a & b \\ \hline
2 & 0 & -81.4 & -75.6 & -102 & -79 & -72.3 & -70 \\
4 & 0 & -82.0 & -94.3 & -121 & -149 & -128.8 & -140.6 \\
4 & 3 & 2362 & 2841 & 3698 & 4044 & 4115 & 4191 \\
6 & 0 & 11.26 & 12.75 & 14.06 & 22.62 & 14.95 & 15.1 \\
6 & 3 & 153 & 216 & 342 & 347 & 164 & 279 \\
6 & 6 & 106 & 150 & 187 & 37 & 229 & 101 \\
&  &  &  &  &  &  &
\end{tabular}
\end{center}
\end{table}


\begin{table}[h]
\caption{Structure of the symmetric dimer Pr$^{3+}$-Cd$^{2+}$ vacancy-Pr$
^{3+}$ (lattice constants $a=0.7675$~nm; $c=0.6722~nm$; $\nu=0.1656$
\protect\cite{R19}).}
\begin{center}
\begin{tabular}{l|lll}
  & \multicolumn{3}{c}{Coordinates} \\
Ion & $x/a$ & $y/a$ & $z/c$ \\ \hline
(Pr$^{3+}$)$_{1,2}$ & 0 & 0 & $\pm[0.5-0.06003]$ \\
(Br$^-$)$_{1,4}$ & $\pm[\sqrt{3}\nu-0.01032]$ & 0 & $\pm[0.75-0.01400]$ \\
(Br$^-$)$_{2,5}$ & $\pm[-\sqrt{3}\nu+0.01032]/2$ & $\pm\sqrt{3}[\sqrt{3}
\nu-0.01032]/2$ & $\pm[0.75-0.01400]$ \\
(Br$^-$)$_{3,6}$ & $\pm[-\sqrt{3}\nu+0.01032]/2$ & $\pm\sqrt{3}[-\sqrt{3}
\nu+0.01032]/2$ & $\pm[0.75-0.01400]$ \\
(Br$^-$)$_{7,10}$ & $\pm[-\sqrt{3}\nu-0.03022]$ & 0 & $\pm[0.25-0.00152]$ \\
(Br$^-$)$_{8,11}$ & $\pm[\sqrt{3}\nu+0.03022]/2$ & $\pm\sqrt{3}[-\sqrt{3}
\nu-0.03022]/2$ & $\pm[0.25-0.00152]$ \\
(Br$^-$)$_{9,12}$ & $\pm[\sqrt{3}\nu+0.03022]/2$ & $\pm\sqrt{3}[\sqrt{3}
\nu+0.03022]/2$ & $\pm[0.25-0.00152]$ \\
(Cs$^+$)$_{1,4}$ & $\pm[-1/\sqrt{3}-0.02051]$ & 0 & $\pm[0.75+0.00469]$ \\
(Cs$^+$)$_{2,5}$ & $\pm[1/\sqrt{3}+0.02051]/2$ & $\pm\sqrt{3}[-1/\sqrt{3}
-0.02051]/2$ & $\pm[0.75+0.00469]$ \\
(Cs$^+$)$_{3,6}$ & $\pm[1/\sqrt{3}+0.02051]/2$ & $\pm\sqrt{3}[1/\sqrt{3}
+0.02051]/2$ & $\pm[0.75+0.00469]$ \\
(Cs$^+$)$_{7,10}$ & $\pm[1/\sqrt{3}-0.00602]$ & 0 & $\pm[0.25-0.01938]$ \\
(Cs$^+$)$_{8,11}$ & $\pm[-1/\sqrt{3}+0.00602]/2$ & $\pm\sqrt{3}[1/\sqrt{3}
-0.00602]/2$ & $\pm[0.25-0.01938]$ \\
(Cs$^+$)$_{9,12}$ & $\pm[-1/\sqrt{3}+0.00602]/2$ & $\pm\sqrt{3}[-1/\sqrt{3}
+0.00602]/2$ & $\pm[0.25-0.01938]$ \\
(Cd$^{2+}$)$_{1,2}$ & 0 & 0 & $\pm[1-0.00290]$ \\
(Br$^-$)$_{13-16}$ & $\pm[-\sqrt{3}(\nu-0.5)+0.00404]$ & $\pm[
\pm(0.5+0.00532)]$ & $\pm[0.25+0.00231]$ \\
(Br$^-$)$_{17-20}$ & $\pm[0.5 \sqrt{3}\nu+0.00258]$ & $\pm[
\pm(1.5\nu-1-0.00616]$ & $\pm[0.25+0.00231]$ \\
(Br$^-$)$_{21-24}$ & $\pm[0.5 \sqrt{3}(\nu-1)-0.00662]$ & $\pm[
\pm(1.5\nu-0.5-0.00084)]$ & $\pm[0.25+0.00231]$ \\
(Br$^-$)$_{25-28}$ & $\pm[\sqrt{3}(\nu-0.5)+0.00151]$ & $\pm[
\pm(0.5-0.00288) ]$ & $\pm[0.75+0.00058]$ \\
(Br$^-$)$_{29,32}$ & $\pm[-0.5 \sqrt{3}\nu+0.00174]$ & $\pm[
\pm(1.5\nu-1+0.00275)]$ & $\pm[0.75+0.00058]$ \\
(Br$^-$)$_{33-36}$ & $\pm[0.5\sqrt{3}(1-\nu)-0.00325]$ & $\pm[
\pm(1.5\nu-0.5-0.00013)]$ & $\pm[0.75+0.00058]$ \\
&  &  &
\end{tabular}
\end{center}
\end{table}

\end{document}